\documentclass[aps,showpacs,tightenlines,twocolumn,nofootinbib,nobibnotes,superscriptaddress]{revtex4-1}
\usepackage{amsmath,amssymb,amsfonts,bm}
\usepackage{graphicx}
\usepackage{epstopdf}
\usepackage{dcolumn}
\usepackage{mathrsfs}
\usepackage{tgtermes}
\usepackage[colorlinks=true,linkcolor=red,citecolor=blue, urlcolor=blue,bookmarks=false]{hyperref}
\bibpunct{[}{]}{,}{n}{}{}
\usepackage{verbatim}
\usepackage{amsthm,amsmath,amssymb}
\usepackage{mathrsfs}
\usepackage{booktabs}
\usepackage{multirow}

\begin{document}
\title{Periodically driven four-dimensional topological insulator with tunable second Chern number}
\date{\today}
\author{Zheng-Rong Liu}
\affiliation{Department of Physics, Hubei University, Wuhan 430062, China}
\author{Rui Chen}\email{chenr@hubu.edu.cn}
\affiliation{Department of Physics, Hubei University, Wuhan 430062, China}
\author{Bin Zhou}\email{binzhou@hubu.edu.cn}
\affiliation{Department of Physics, Hubei University, Wuhan 430062, China}
\affiliation{Key Laboratory of Intelligent Sensing System and Security of Ministry of Education, Hubei University, Wuhan 430062, China}

\begin{abstract}
In recent years, Floquet engineering has attracted considerable attention as a promising approach for tuning topological phase transitions. In this work, we investigate the effects of high-frequency time-periodic driving in a four-dimensional (4D) topological insulator, focusing on topological phase transitions at the off-resonant quasienergy gap. The 4D topological insulator hosts gapless three-dimensional boundary states characterized by the second Chern number $C_{2}$. We demonstrate that the second Chern number of 4D topological insulators can be modulated by tuning the amplitude of time-periodic driving. This includes transitions from a topological phase with $C_{2}=\pm3$ to another topological phase with $C_{2}=\pm1$, or to a topological phase with an even second Chern number $C_{2}=\pm2$ which is absent in the 4D static system. Finally, the approximation theory in the high-frequency limit further confirms the numerical conclusions.
\end{abstract}

\maketitle

\section{Introduction}
The discovery of the quantum Hall effect (QHE) opens the door to the field of topological matter~\cite{PhysRevLett.45.494}. In recent years,  tremendous effort has been devoted to realizing the QHE in higher dimensions~\cite{PhysRevLett.119.136806, 10.1038/s41586-018-0798-3, 10.1038/s41586-019-1180-9, PhysRevLett.127.066801, PhysRevLett.125.206601, PhysRevLett.56.85, PhysRevLett.63.1988, PhysRevB.41.11417, PhysRevB.45.13488, PhysRevB.58.10778, PhysRevLett.86.1062, PhysRevLett.99.146804, doi:10.1126/sciadv.1501117, PhysRevB.98.081101, PhysRevLett.125.036602, PhysRevB.102.075304, PhysRevB.101.161201, 10.1038/s41535-021-00399-2}. In 2001, Zhang and Hu proposed a four-dimensional (4D) generalization of the QHE~\cite{Zhang_2001}. The 4D QHE is characterized by the second Chern number, a topological invariant that describes the quantized nonlinear response of current to electric and magnetic fields~\cite{PhysRevB.78.195424, Mochol_Grzelak_2018, doi:10.1126/science.aam9031}. The 4D topological insulator (TI) exhibiting the 4D QHE supports fully gapped bulk and gapless three-dimensional (3D) boundary states~\cite{PhysRevB.78.195424, PhysRevLett.111.186803, PhysRevLett.129.196602, PhysRevResearch.2.023364, PhysRevB.108.085306}. The 4D TI is impossible to realize in real materials due to the limitation of spatial dimensions. In engineered systems, the recent proposals for realizing the 4D TIs include: introducing synthetic dimensions in two-dimensional (2D) or 3D systems~\cite{PhysRevLett.115.195303, PhysRevA.93.043827, PhysRevX.11.011016, 10.1093/nsr/nwac289}, mapping high-dimensional models onto lower-dimensional systems~\cite{10.1038/nature25000, 10.1038/nature25011, PhysRevLett.111.226401, PhysRevLett.109.135701, PhysRevB.98.094434, PhysRevB.98.125431}, and implementing 4D lattices by constructing appropriate capacitive and inductive connections in electric circuits~\cite{10.1038/s41467-020-15940-3, 10.1093/nsr/nwaa065, 10.1038/s41467-023-36359-6, 10.1038/s41467-023-36767-8}. Experimentally, 4D topological states have been realized in acoustic lattices~\cite{PhysRevX.11.011016}, photonic crystals~\cite{10.1038/nature25011, 10.1093/nsr/nwac289, 10.1515/nanoph-2022-0778}, an angled optical superlattice with ultracold bosonic atoms~\cite{10.1038/nature25000}, and electric circuits~\cite{10.1038/s41467-020-15940-3, 10.1093/nsr/nwaa065, 10.1038/s41467-023-36359-6, 10.1038/s41467-023-36767-8}.

Floquet engineering, which applies time-periodic external fields to manipulate quantum systems, has emerged as a promising route to modulate topological phases~\cite{PhysRevLett.114.246802, PhysRevB.96.020507, 10.1038/nmat4156, 10.1038/nphys1926, doi:10.1126/science.1239834, 10.1038/nphys3609, 10.1038/s41567-019-0698-y, 10.1038/s41586-022-05610-3, 10.1038/s41467-019-12231-4, 10.1038/s41586-023-05850-x, PhysRevLett.110.016802, 10.1038/s41467-023-42139-z, PhysRevB.97.155152, PhysRevB.98.235159}. It was found that time-periodic perturbations can drive trivial systems out of equilibrium to produce topological boundary states, and such topologically nontrivial phases driven by periodic external fields are called Floquet topological insulators (FTI)~\cite{PhysRevB.79.081406, PhysRevB.82.235114, 10.1038/nphys1926, PhysRevLett.107.216601, 10.1038/nature12066, 10.1002/pssr.201206451, 10.1038/s42254-020-0170-z}. Recently, the research on periodically driven Floquet topological matter in solid-state materials~\cite{10.1038/nphys1926, doi:10.1126/science.1239834, 10.1038/s41567-019-0698-y, 10.1038/s41586-022-05610-3, PhysRevB.79.081406, PhysRevB.89.121401, PhysRevB.90.115423, PhysRevLett.113.236803, PhysRevLett.113.266801, PhysRevLett.114.056801, 10.1038/nphys3609, wan2023photoinduced}, photonic crystals~\cite{10.1038/s41467-019-12231-4, 10.1038/nature12066, 10.1038/ncomms13756, PhysRevLett.122.173901, PhysRevB.107.174313, 10.1038/ncomms1872, 10.1038/ncomms13918, PhysRevLett.124.253601, 10.1038/s41563-020-0641-8}, acoustic lattices~\cite{10.1038/ncomms11744, 10.1038/ncomms13368, 10.1038/s41467-021-27552-6, PhysRevLett.129.254301}, electric circuits~\cite{dabiri2023electric}, and cold atom systems~\cite{10.1038/nature13915, 10.1038/s41567-019-0417-8, 10.1038/s41567-020-0949-y, PhysRevLett.130.043201} has attracted considerable attention. In 2018, Peng and Refael presented a one-dimensional (1D) model with three quasiperiodic drives, realizing a 4D QHE that allows a bulk energy conversion by treating the three drives as synthetic dimensions~\cite{PhysRevB.97.134303}. Motivated by the above mentioned researches, a question naturally arises whether time-periodic driving can modulate topological phase transitions in 4D TIs.

In this work, we present a systematic study on the time-periodically driven 4D TI with a high-frequency time-periodic driving. Here, we only focus on the high-frequency case where the frequency of the time-periodic driving is larger than the bandwidth of the static system, and there is no overlap between the undriven quasienergy bands and the driven quasienergy bands. We consider a 4D time-periodic vector potential ${\boldsymbol{V}}(V_{x}, V_{y}, V_{z}, V_{w})$ as the time-periodic driving. When $V_{x}\neq0$ or $V=V_{n}\neq0$ ($n=x,y,z,w$), the second Chern number of the 4D system can be changed by tuning the amplitude of the time-periodic driving. However, there is no newly formed topological phase with distinct second Chern number. Furthermore, when $V_{x}=V_{y}\neq0$ or $V_{x}=V_{y}=V_{z}\neq0$, one can find that the time-periodic driving transforms a topological phase with $C_{2}=\pm3$ to another topological phase with $C_{2}=\pm1$, or to a topological phase with an even second Chern number $C_{2}=\pm2$ which is absent in the 4D static system. By solving for the effective Hamiltonian in the high-frequency limit, we present analytical expressions representing the closure of the bulk gap. We find that the transition points in the phase diagram fit well with the bulk gap closure points.

The rest of the paper is organized as follows. In Sec.~\ref{SecII.A}, we introduce the 4D Dirac model describing the 4D TI, and we present the method for calculating the second Chern number in Sec.~\ref{SecII.B}. In Sec.~\ref{SecIII}, we investigate the 4D TI driven by time-periodic driving. In Sec.~\ref{SecIII.A}, we introduce time-periodic driving in the 4D Dirac model and transform the time-dependent Hamiltonian into the time-independent Floquet Hamiltonian based on the Floquet theory. In Sec.~\ref{SecIII.B}, we explore the effect of the time-periodic driving ${\boldsymbol{V}}(V_{x}, 0, 0, 0)$ in the 4D TI. Then, we investigate the effect of the time-periodic driving ${\boldsymbol{V}}(V_{x}, V_{y}, 0, 0)$ in the 4D TI in Sec.~\ref{SecIII.C}. Finally, we summarize our conclusions in Sec.~\ref{Conclusion}.

\section{Static system}
\label{SecII}

\subsection{Static model}
\label{SecII.A}
\begin{figure}[t]
	\includegraphics[width=8.5cm]{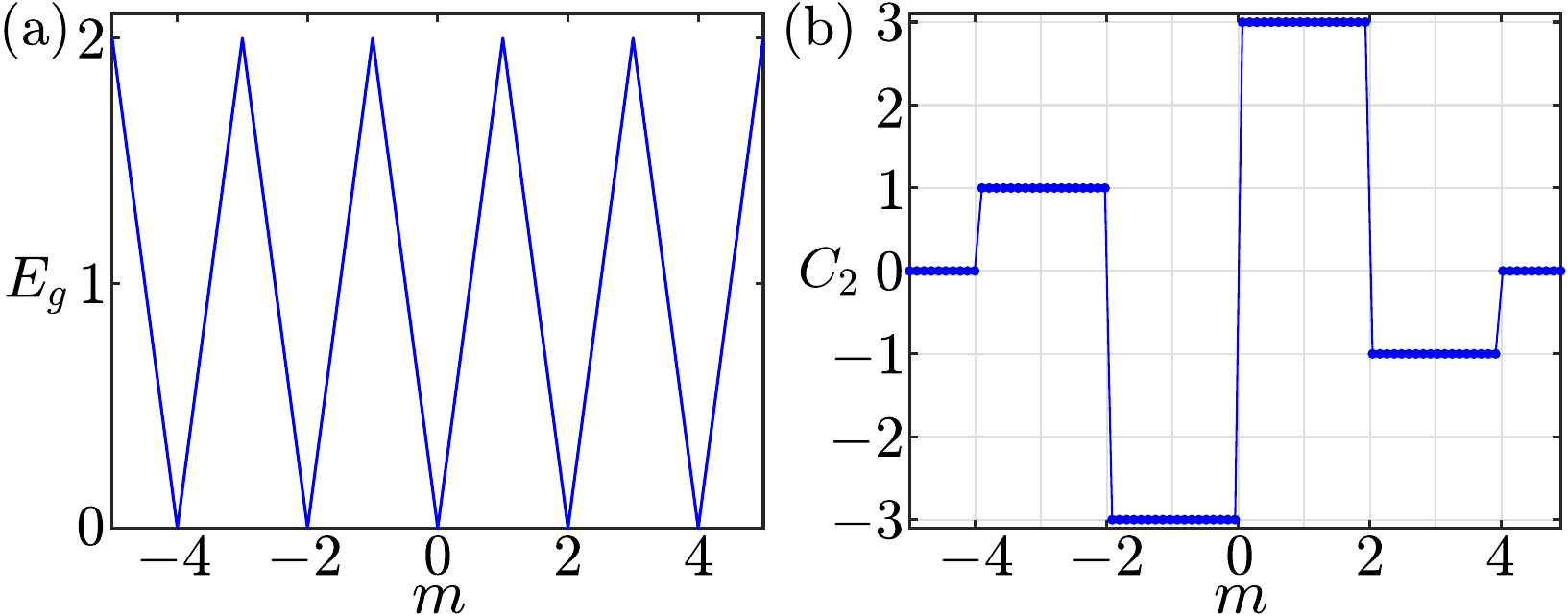} \caption{(a) The bulk gap $E_{g}$ as a function of the parameter $m$. (b) The second Chern number $C_{2}$ as a function of the parameter $m$.}%
	\label{fig1}
\end{figure}

The Dirac model describing the 4D TI is given by the following equation~\cite{PhysRevB.78.195424}:
\begin{align}
H(\textbf{k})=&\sin(k_{x})\Gamma_{2}+\sin(k_{y})\Gamma_{3}+\sin(k_{z})\Gamma_{4}+\sin(k_{w})\Gamma_{5}\nonumber\\
&+m(\textbf{k})\Gamma_{1},
\end{align}
where the Dirac matrices $\Gamma_{j}=(s_{x}\otimes s_{0}, s_{y}\otimes s_{0}, s_{z}\otimes s_{x}, s_{z}\otimes s_{y}, s_{z}\otimes s_{z}), j=1, 2, 3, 4, 5,$ satisfying the anticommutation relations $\{\Gamma_{i},\Gamma_{j}\}=2\delta_{ij}$. $m(\textbf{k})=m+c[\cos(k_{x})+\cos(k_{y})+\cos(k_{z})+\cos(k_{w})]$, where $m$ is the Dirac mass and $c$ denotes the nearest-neighbour hopping amplitude. In subsequent calculations, $c=1$. The doubly degenerate eigenvalues of $H(\textbf{k})$ are
\begin{align}
E(\textbf{k})=\pm\sqrt{4+2 m(\textbf{k})^{2}-\sum_{n=x,y,z,w}\cos(2 k_{n})}/\sqrt{2},
\end{align}
and the bandwidth of the system is $E_{W}=2|m|+8$. In Fig.~\ref{fig1}(a), we plot the bulk gap $E_{g}$ with respect to the Dirac mass $m$. Remarkably, the bulk gap $E_{g}$ is closed at $m=\pm4,m=\pm2,m=0$.

\label{SecII.B}

\subsection{Second Chern number}
\begin{figure}[t]
	\includegraphics[width=8.5cm]{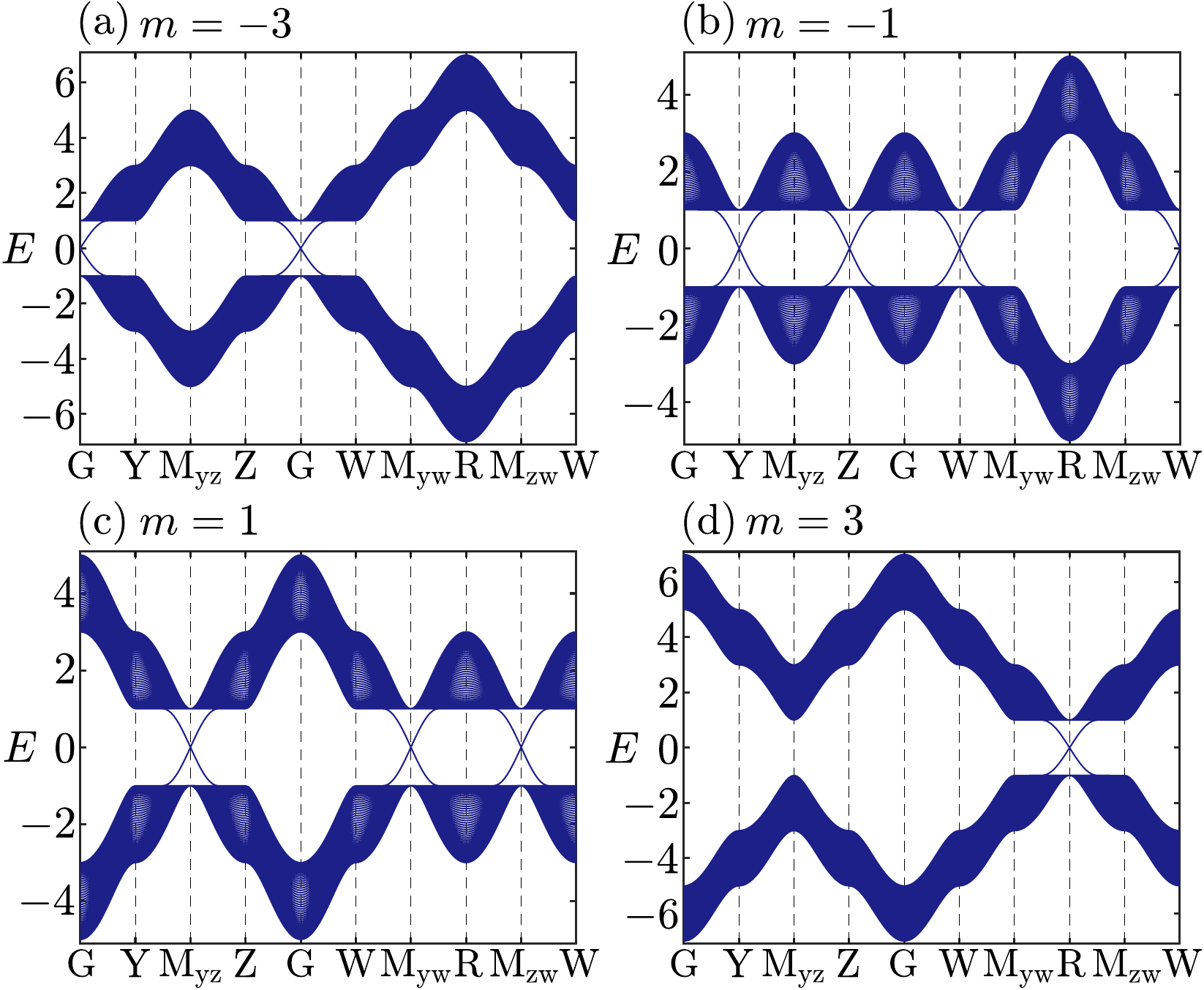} \caption{Energy spectra for (a) $m=-3$, (b) $m=-1$, (c) $m=1$, and (d) $m=3$ when the OBC along the $x$ direction. The horizontal axis shows the high symmetry points in the FBZ of the quasi-3D system, ${\rm{G}}=(k_{y}=0, k_{z}=0, k_{w}=0)$, $\rm{Y}=(\pi, 0, 0)$, $\rm{Z}=(0, \pi, 0)$, $\rm{W}=(0, 0, \pi)$, $\rm{M_{yz}}=(\pi, \pi, 0)$, $\rm{M_{yw}}=(\pi, 0, \pi)$, $\rm{M_{zw}}=(0, \pi, \pi)$, $\rm{R}=(\pi, \pi, \pi)$.}%
	\label{fig2}
\end{figure}

The second Chern number is used to describe the nonlinear response of the current to an electric field and a magnetic field in 4D system~\cite{Zhang_2001}. The second Chern number is given by the following formula~\cite{PhysRevB.78.195424, Mochol_Grzelak_2018, doi:10.1126/science.aam9031, PRXQuantum.2.010310, PhysRevB.108.085114}:
\begin{align}
C_{2}=\frac{1}{4\pi^{2}}\int_{\rm{FBZ}}d\textbf{k}\text{Tr}[F_{xy}F_{zw}+F_{wx}F_{zy}+F_{zx}F_{yw}],
\end{align}
with the non-Abelian Berry curvature
\begin{align}
F_{mn}^{\alpha\beta}=\partial_{m}A_{n}^{\alpha\beta}-\partial_{n}A_{m}^{\alpha\beta}+i[A_{m},A_{n}]^{\alpha\beta},
\end{align}
where $m, n=x, y, z, w$ and the Berry connection of the occupied bands $A_{m}^{\alpha\beta}=-i\left\langle u^{\alpha}(\textbf{k})\right\vert\frac{\partial}{\partial k_{m}}\left\vert u^{\beta}(\textbf{k})\right\rangle$, and $\left\vert u^{\alpha}(\textbf{k})\right\rangle$ denotes the occupied eigenstates with $\alpha=1, \dots, N_{\rm{occ}}$. Figure \ref{fig1}(b) shows the second Chern number $C_{2}$ as a function of the Dirac mass $m$. It is obvious that the phase transition of the system is accompanied by the closure of the bulk gap. Qi~$et~al$. demonstrated that the number of 3D gapless boundary states with linear dispersion in the 4D system is equal to the value of the second Chern number~\cite{PhysRevB.78.195424}. In the region $-4<m<-2$, the second Chern number $C_{2}=1$. In the first Brillouin zone (FBZ) of the quasi-3D system [open boundary conditions (OBC) along the $x$ direction], there is a single gapless Dirac cone cross at the point ${\rm{G}}=(k_{y}=0, k_{z}=0, k_{w}=0)$ as shown in Fig.~\ref{fig2}(a). In the region $-2<m<0$, the second Chern number $C_{2}=-3$. There are three gapless Dirac cones cross at the points $\rm{Y}=(\pi, 0, 0)$, $\rm{Z}=(0, \pi, 0)$, and $\rm{W}=(0, 0, \pi)$ [Fig.~\ref{fig2}(b)]. The results for $m>0$ are symmetrically distributed with those for $m<0$. In the region $0<m<2$, the second Chern number $C_{2}=3$. There are three gapless Dirac cones cross at the points $\rm{M_{yz}}=(\pi, \pi, 0)$, $\rm{M_{yw}}=(\pi, 0, \pi)$, and $\rm{M_{zw}}=(0, \pi, \pi)$ [Fig.~\ref{fig2}(c)]. Finally, in the region $2<m<4$, the second Chern number $C_{2}=-1$. There is one gapless Dirac cone cross at the vertex of the FBZ $\rm{R}=(\pi, \pi, \pi)$ [Fig.~\ref{fig2}(d)].

\section{Floquet system}
\label{SecIII}

\subsection{Time-periodically driven model and Floquet Hamiltonian}
\label{SecIII.A}
The Floquet theory is often used to study time-periodic systems~\cite{PhysRevB.96.054207}. For a time-periodic Hamiltonian $H(\tau)=H(\tau+T)$ with the period $T=2\pi/\omega$ ($\omega$ is the frequency), the wave function $\Psi_{\alpha}(\tau)=e^{-i\varepsilon_{\alpha}\tau}\psi_{\alpha}$ of the Schr\"{o}dinger equation $i\partial_{\tau}\Psi(\tau)=H(\tau)\Psi(\tau)$ can be obtained by employing the Floquet theory, where $\varepsilon_{\alpha}$ is the $\alpha$-th quasienergy. Using the Fourier transformation, the time-dependent Schr\"{o}dinger equation can be converted into a time-independent Schr\"{o}dinger equation:
\begin{align}
H_{F}\psi_{\alpha}=\varepsilon_{\alpha}\psi_{\alpha},
\end{align}
where
\begin{align}
H_{F,qp}&=q\omega\delta_{qp}I+H_{p-q},\\
H_{0}&=\frac{1}{T}\int_{0}^{T}d\tau H(\tau),\nonumber\\
H_{p-q}&=\frac{1}{T}\int_{0}^{T}d\tau H(\tau)e^{i (p-q)\omega\tau},\nonumber
\end{align}
$I$ is an identity matrix of the same size as $H(\tau)$. $p, q$ take $0, \pm1, \pm2, ...$ and the Floquet Hamiltonian $H_{F}$ is an infinite Hamiltonian. In the numerical calculations, $p$ and $q$ are taken to be finite integers until the results converge. When $\omega > E_{W}$, there is no overlap between quasienergy bands at different intervals $\varepsilon\in[(n-1/2)\omega,(n+1/2)\omega]$, and $n$ is an integer. In the subsequent content, we restrict our attention to the quasienergy interval $\varepsilon\in[-\omega/2,\omega/2]$.

We apply the time-periodic vector potential ${\boldsymbol{V}}(V_{x}, V_{y}, V_{z}, V_{w})$ to the 4D Dirac model $H(\textbf{k})$, then the time-dependent Hamiltonian $H(\textbf{k}, \tau)$ is given by the following expression:
\begin{align}
H(\textbf{k}, \tau)&=\sum_{j=1}^{5}h_{j}(\textbf{k}, \tau)\Gamma_{j},
\end{align}
with
\begin{align}
h_{1}(\textbf{k}, \tau)&=m+c\sum_{n=x, y, z, w}\cos[k_{n}+V_{n}\cos(\omega\tau)],\nonumber\\
h_{2}(\textbf{k}, \tau)&=\sin[k_{x}+V_{x}\cos(\omega\tau)],\nonumber\\
h_{3}(\textbf{k}, \tau)&=\sin[k_{y}+V_{y}\cos(\omega\tau)],\nonumber\\
h_{4}(\textbf{k}, \tau)&=\sin[k_{z}+V_{z}\cos(\omega\tau)],\nonumber\\
h_{5}(\textbf{k}, \tau)&=\sin[k_{w}+V_{w}\cos(\omega\tau)],
\end{align}
where $V_{n}$ ($n=x, y, z, w$) is the amplitude of the time-periodic vector potential. After the Fourier transformation, the diagonal block matrices in the Floquet Hamiltonian $H_{F}$ are shown below:
\begin{align}
H_{0}&=\sum_{j=1}^{5}h_{0, j}\Gamma_{j},
\end{align}
with
\begin{align}
h_{0, 1}&=m+c\sum_{n=x, y, z, w}\cos(k_{n})\mathcal{J}_{0}(V_{n}),\nonumber\\
h_{0, 2}&=\sin(k_{x})\mathcal{J}_{0}(V_{x}),
h_{0, 3}=\sin(k_{y})\mathcal{J}_{0}(V_{y}),\nonumber\\
h_{0, 4}&=\sin(k_{z})\mathcal{J}_{0}(V_{z}),
h_{0, 5}=\sin(k_{w})\mathcal{J}_{0}(V_{w}),
\end{align}
and the off-diagonal block matrices
\begin{align}
H_{-l}&=\sum_{j=1}^{5}h_{-l, j}\Gamma_{j}, H_{+l}=H_{-l}^{\dagger},
\end{align}
with
\begin{align}
h_{-l, 1}&=\frac{c}{2}\sum_{n=x, y, z, w}\left[e^{i k_{n}}+(-1)^{l}e^{-i k_{n}}\right]\mathcal{J}_{l}(V_{n}),\nonumber\\
h_{-l, 2}&=-\frac{i}{2}\left[e^{i k_{x}}-(-1)^{l}e^{-i k_{x}}\right]\mathcal{J}_{l}(V_{x}),\nonumber\\
h_{-l, 3}&=-\frac{i}{2}\left[e^{i k_{y}}-(-1)^{l}e^{-i k_{y}}\right]\mathcal{J}_{l}(V_{y}),\nonumber\\
h_{-l, 4}&=-\frac{i}{2}\left[e^{i k_{z}}-(-1)^{l}e^{-i k_{z}}\right]\mathcal{J}_{l}(V_{z}),\nonumber\\
h_{-l, 5}&=-\frac{i}{2}\left[e^{i k_{w}}-(-1)^{l}e^{-i k_{w}}\right]\mathcal{J}_{l}(V_{w}),
\end{align}
where $\mathcal{J}_{l}(V_{n})$ is the $l$-th Bessel function of the first kind, $l=|p-q|$ ($p\neq q$). Then, the Floquet Hamiltonian $H_{F}$ is given by the following matrix:
\begin{align}
H_{F}=\begin{pmatrix}
\ddots & \vdots & \vdots & \vdots & \ddots \\
\cdots & H_{0}-\omega & H_{+1} & H_{+2} & \cdots \\
\cdots & H_{-1} & H_{0} & H_{+1} & \cdots \\
\cdots & H_{-2} & H_{-1} & H_{0}+\omega  & \cdots \\
\ddots & \vdots & \vdots & \vdots & \ddots
\end{pmatrix}.
\end{align}

\subsection{Time-periodic driving ${\boldsymbol{V}}(V_{x}, 0, 0, 0)$}
\label{SecIII.B}
\begin{figure}[t]
	\includegraphics[width=8.5cm]{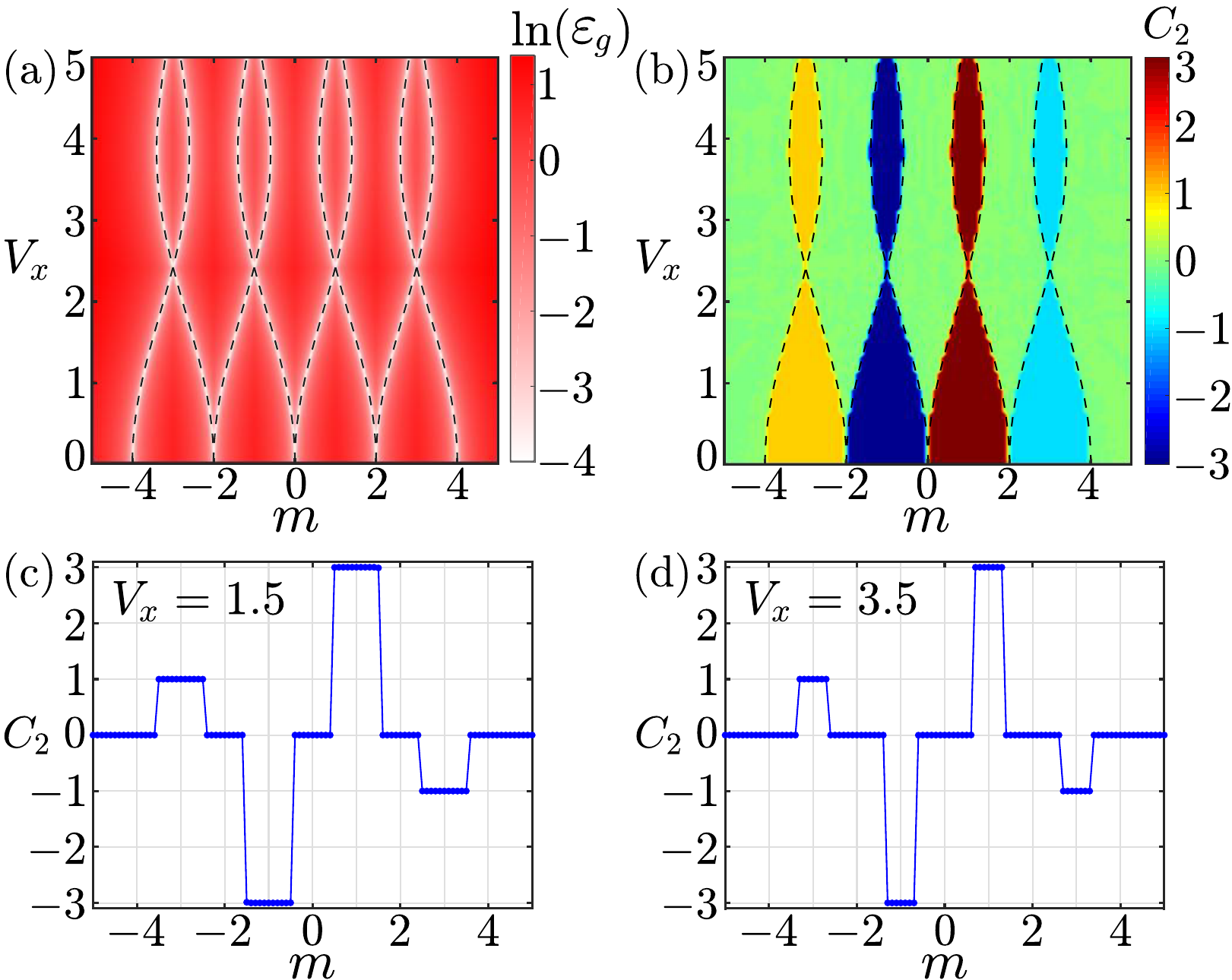} \caption{(a) The logarithm of the bulk gap ${\rm{ln}}(\varepsilon_{g})$ as a function of $m$ and $V_{x}$. The white area indicates that the bulk gap is close to zero, and black dashed lines are given by $m=\pm3\pm\mathcal{J}_{0}(V_{x})$, $m=\pm1\pm\mathcal{J}_{0}(V_{x})$. (b) Second Chern number $C_{2}$ as a function of $m$ and $V_{x}$. The second Chern number $C_{2}$ as a function of the parameter $m$ for (c) $V_{x}=1.5$ and (d) $V_{x}=3.5$.}%
	\label{fig3}
\end{figure}

\begin{figure}[t]
	\includegraphics[width=8.5cm]{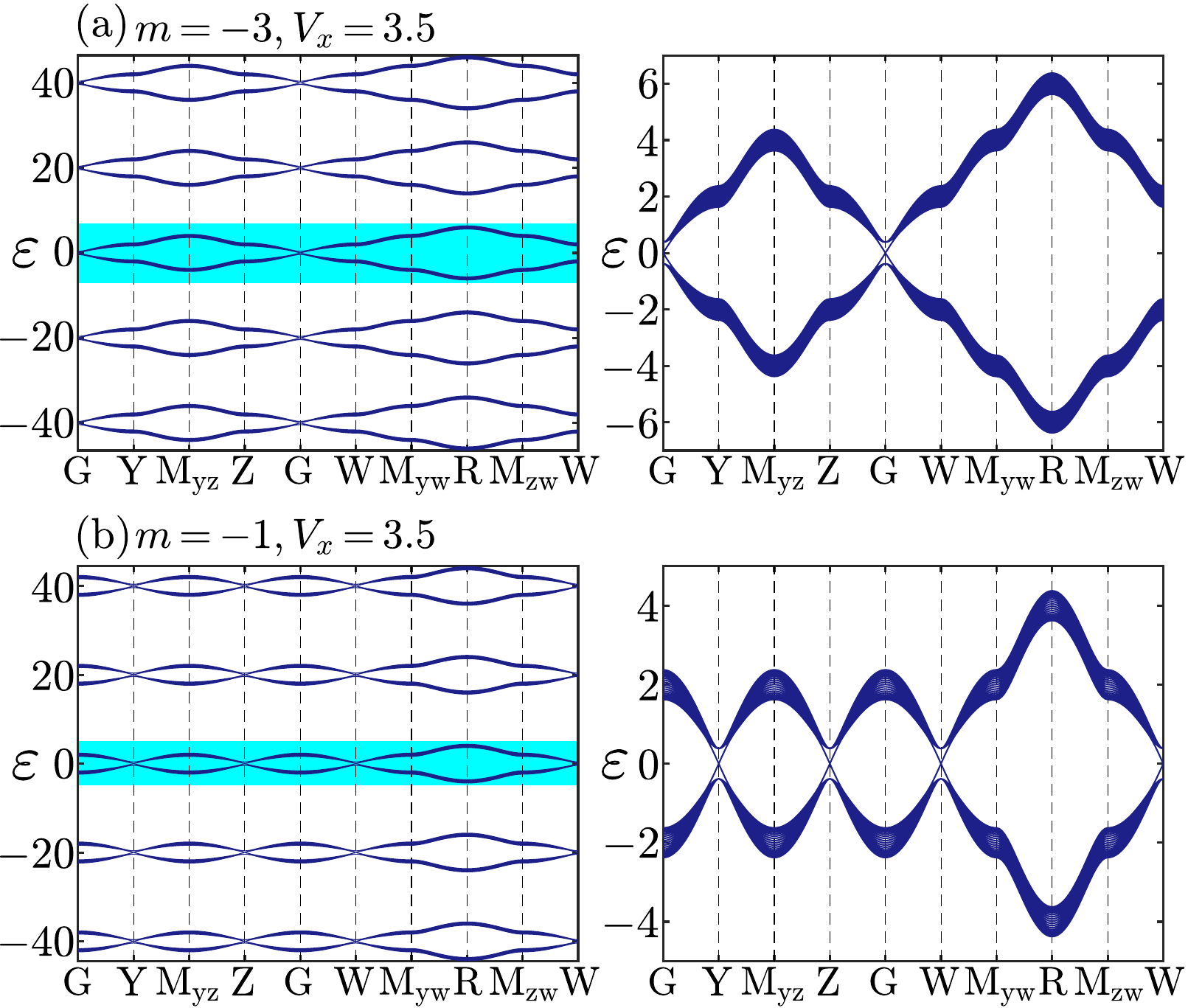} \caption{Quasienergy spectra for (a) $m=-3$ and (b) $m=-1$ when the OBC along the $x$ direction. The horizontal axis shows the high symmetry points in the FBZ of the quasi-3D system, ${\rm{G}}=(k_{y}=0, k_{z}=0, k_{w}=0)$, $\rm{Y}=(\pi, 0, 0)$, $\rm{Z}=(0, \pi, 0)$, $\rm{W}=(0, 0, \pi)$, $\rm{M}_{yz}=(\pi, \pi, 0)$, $\rm{M_{yw}}=(\pi, 0, \pi)$, $\rm{M_{zw}}=(0, \pi, \pi)$, $\rm{R}=(\pi, \pi, \pi)$. The right panels show the cyan area of the left panels. The amplitude and frequency of the periodic driving are $V_{x}=3.5$ and $\omega=20$.}%
	\label{fig4}
\end{figure}

In this subsection, we first consider the time-periodic driving ${\boldsymbol{V}}(V_{x}, 0, 0, 0)$. Then, the diagonal block matrices in the Floquet Hamiltonian $H_{F}$ are given by:
\begin{align}
H_{0}=&\sum_{j=1}^{5}h_{0, j}\Gamma_{j},\label{Eq10}
\end{align}
with
\begin{align}
h_{0, 1}=&m+c[ \cos(k_{x})\mathcal{J}_{0}(V_{x})\nonumber\\&+\cos(k_{y})+\cos(k_{z})+\cos(k_{w}) ],\nonumber\\
h_{0, 2}=&\sin(k_{x})\mathcal{J}_{0}(V_{x}),
h_{0, 3}=\sin(k_{y}),\nonumber\\
h_{0, 4}=&\sin(k_{z}),
h_{0, 5}=\sin(k_{w}),
\end{align}
and the off-diagonal block matrices
\begin{align}
H_{-l}=&\sum_{j=1}^{5}h_{-l, j}\Gamma_{j}, H_{+l}=H_{-l}^{\dagger},
\end{align}
with
\begin{align}
h_{-l, 1}=&\frac{c}{2}\left[e^{i k_{x}}+(-1)^{l}e^{-i k_{x}}\right]\mathcal{J}_{l}(V_{x}),\nonumber\\
h_{-l, 2}=&-\frac{i}{2}\left[e^{i k_{x}}-(-1)^{l}e^{-i k_{x}}\right]\mathcal{J}_{l}(V_{x}),\nonumber\\
h_{-l, 3}=&h_{-l, 4}=h_{-l, 5}=0.
\end{align}
By numerically diagonalizing the Floquet Hamiltonian $H_{F}$, we show the logarithm of the bulk gap as a function of $m$ and $V_{x}$ in Fig.~\ref{fig3}(a). Here, the bulk gap refers to the quasienergy gap near $\varepsilon=0$. The color bar converging to white represents the bulk gap close to $0$. In the high-frequency limit (i.e., $\omega\gg E_{W}$), we can obtain the effective Floquet Hamiltonian $H_{\rm{eff}}$. The black dashed lines in Fig.~\ref{fig3}(a) can be given by solving for the effective Hamiltonian, $m=\pm3\pm\mathcal{J}_{0}(V_{x})$, $m=\pm1\pm\mathcal{J}_{0}(V_{x})$ (see Appendix~\ref{AppendixA} for details). Obviously, the results obtained by solving numerically for the Floquet Hamiltonian $H_{F}$ are consistent with those obtained by solving analytically for the effective Hamiltonian $H_{\rm{eff}}$.

Accordingly, we plot the phase diagram of the system as a function of $m$ and $V_{x}$ in Fig.~\ref{fig3}(b). It is found that as the amplitude of the periodic driving $V_{x}$ increases, the topologically nontrivial regions with nonzero second Chern numbers diminish until they completely vanish at $\mathcal{J}_{0}(V_{x,c})=0$. However, when $V_{x}>V_{x,c}\approx2.405$, the topologically nontrivial phase of the system reappears. Figures \ref{fig3}(c) and \ref{fig3}(d) show the second Chern number as a function of the Dirac mass $m$ when $V_{x}<V_{x,c}$ and $V_{x}>V_{x,c}$, respectively. It is found that the reappeared topologically nontrivial phase exhibits the same second Chern number as before the phase transition. Figure~\ref{fig4} illustrates the quasienergy spectra of different topologically nontrivial phases at $V_{x}=3.5$ when the OBC along the $x$ direction. The right panels show the quasienergy spectra in the interval $[-\omega/2,\omega/2]$. The results of the system with OBC along the $y$, $z$, or $w$ direction are the same as the result in Fig.~\ref{fig4}, i.e., the number of gapless 3D topological boundary states is equal to the value of the second Chern number. Here, we only discuss the topological phase transitions for $V_{x}\neq0$. It should be noted that topological phase transitions for $V_{n}\neq0$ ($n=y$, $z$, or $w$) is similar to those presented in this subsection.

\subsection{Topological phases with even second Chern number}
\label{SecIII.C}
\begin{figure}[t]
	\includegraphics[width=8.5cm]{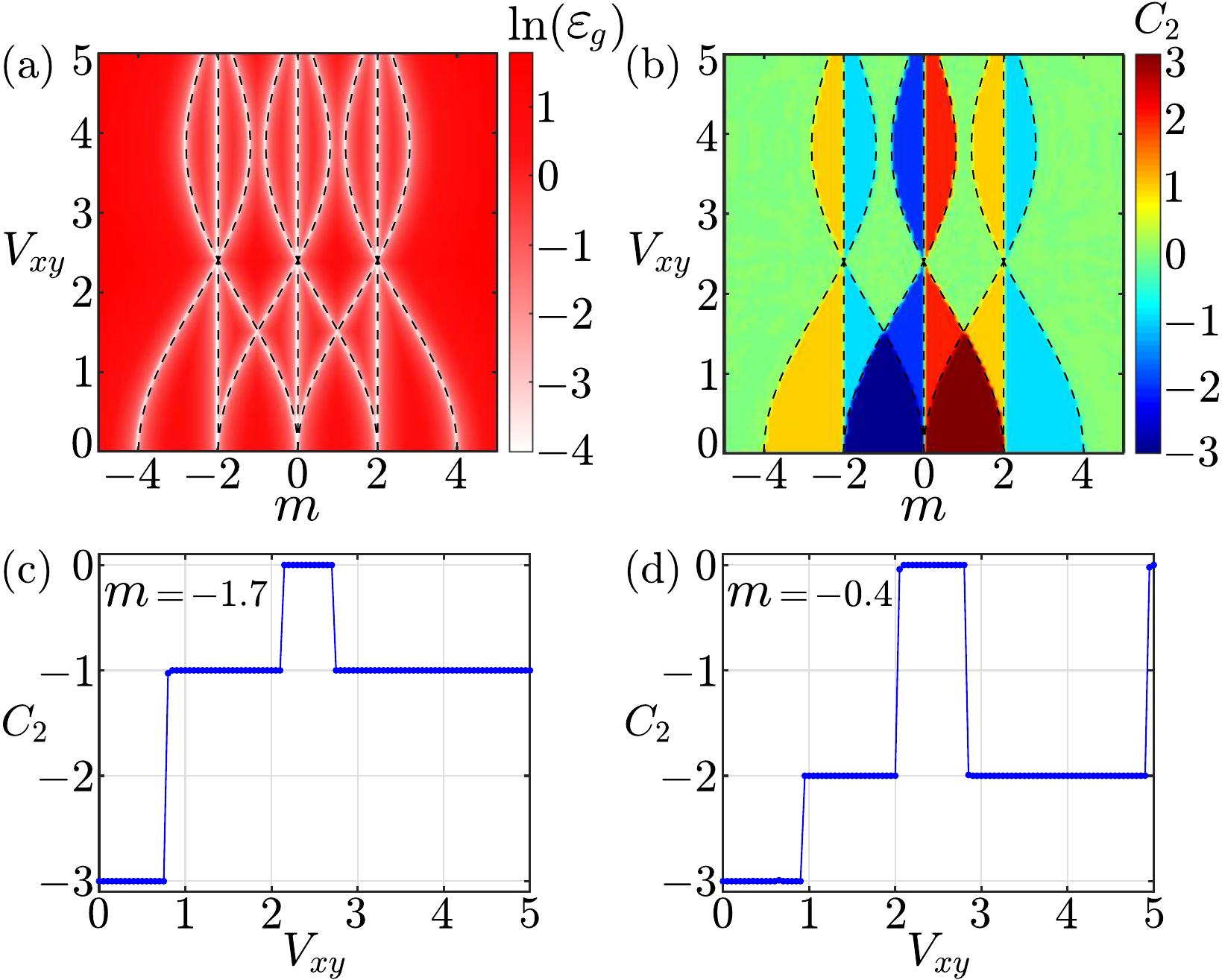} \caption{(a) The logarithm of the bulk gap as a function of $m$ and $V_{xy}$. The white area indicates that the bulk gap is close to zero, and black dashed lines are given by $m=0$, $m=\pm2$, $m=\pm 2\mathcal{J}_{0}(V_{xy})$, and $m=\pm2\pm 2\mathcal{J}_{0}(V_{xy})$. (b) Second Chern number $C_{2}$ as a function of $m$ and $V_{xy}$. The second Chern number $C_{2}$ as a function of $V_{xy}$ for (c) $m=-1.7$ and (d) $m=-0.4$.}%
	\label{fig5}
\end{figure}

\begin{figure*}[t]
	\includegraphics[width=17.5cm]{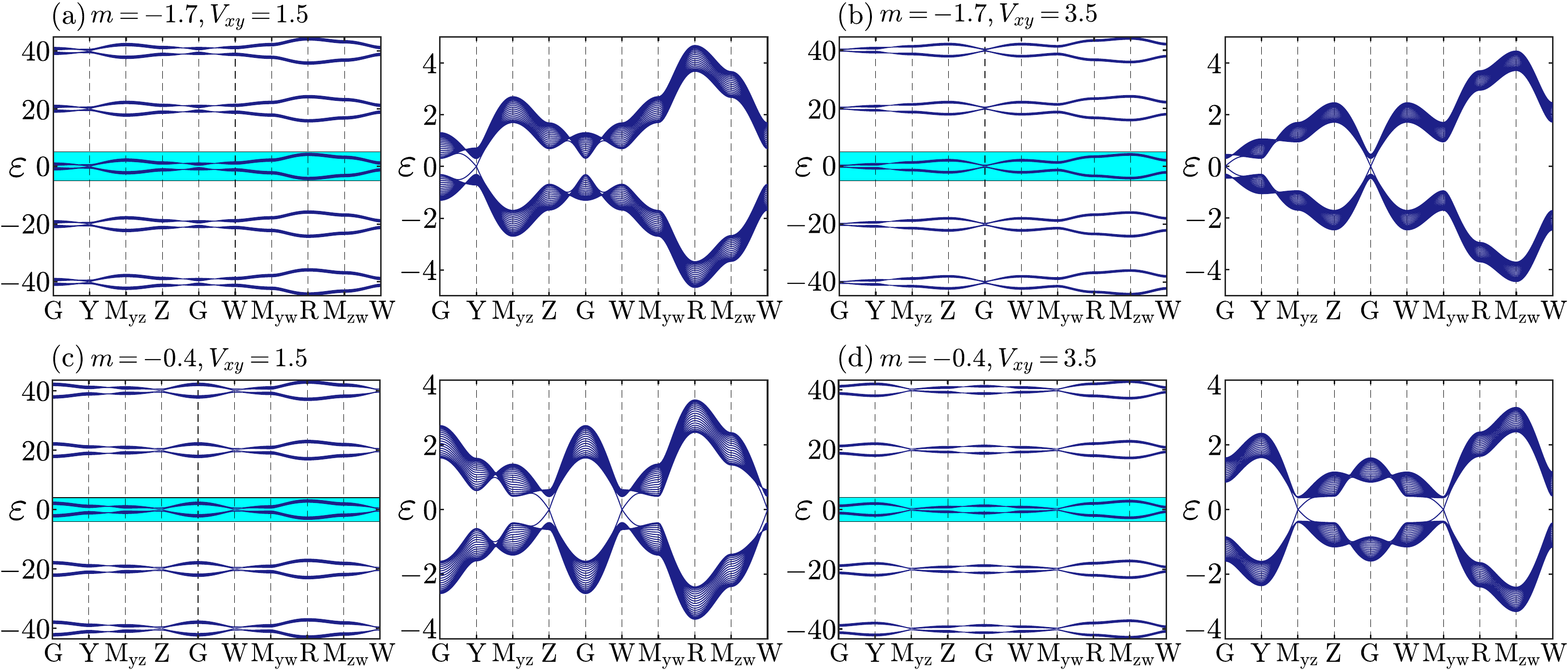} \caption{Quasienergy spectra for the OBC along the $x$ direction. The horizontal axis shows the high symmetry points in the FBZ of the quasi-3D system, ${\rm{G}}=(k_{y}=0, k_{z}=0, k_{w}=0)$, $\rm{Y}=(\pi, 0, 0)$, $\rm{Z}=(0, \pi, 0)$, $\rm{W}=(0, 0, \pi)$, $\rm{M_{yz}}=(\pi, \pi, 0)$, $\rm{M_{yw}}=(\pi, 0, \pi)$, $\rm{M_{zw}}=(0, \pi, \pi)$, $\rm{R}=(\pi, \pi, \pi)$. The right panels show the cyan area of the left panels. The Dirac mass and the amplitude of the periodic driving are $m=-1.7$ and $V_{xy}=1.5$ for (a), $m=-1.7$ and $V_{xy}=3.5$ for (b), $m=-0.4$ and $V_{xy}=1.5$ for (c), $m=-0.4$ and $V_{xy}=3.5$ for (d). Here, we choose $\omega=20$.}%
	\label{fig6}
\end{figure*}

In this subsection, we consider the periodic driving ${\boldsymbol{V}}(V_{x}, V_{y}, 0, 0)$, where $V_{x}=V_{y}=V_{xy}\neq0$. Then, the diagonal block matrices in the Floquet Hamiltonian $H_{F}$ are given by:
\begin{align}
H_{0}=&\sin(k_{x})\mathcal{J}_{0}(V_{xy})\Gamma_{2}+\sin(k_{y})\mathcal{J}_{0}(V_{xy})\Gamma_{3}\nonumber\\
&+\sin(k_{z})\Gamma_{4}+\sin(k_{w})\Gamma_{5}+m(\textbf{k})\Gamma_{1},
\end{align}
and the off-diagonal block matrices
\begin{align}
H_{-l}=&\sum_{j=1}^{5}h_{-l, j}\Gamma_{j}, H_{+l}=H_{-l}^{\dagger},
\end{align}
with
\begin{align}
h_{-l, 1}=&\frac{c}{2}[e^{i k_{x}}+(-1)^{l}e^{-i k_{x}}+e^{i k_{y}}+(-1)^{l}e^{-i k_{y}}]\mathcal{J}_{l}(V_{xy}),\nonumber\\
h_{-l, 2}=&-\frac{i}{2}[e^{i k_{x}}-(-1)^{l}e^{-i k_{x}}]\mathcal{J}_{l}(V_{xy}),\nonumber\\
h_{-l, 3}=&-\frac{i}{2}[e^{i k_{y}}-(-1)^{l}e^{-i k_{y}}]\mathcal{J}_{l}(V_{xy}),\nonumber\\
h_{-l, 4}=&h_{-l, 5}=0,
\end{align}
where $m(\textbf{k})=m+c[\cos(k_{x})\mathcal{J}_{0}(V_{xy})+\cos(k_{y})\mathcal{J}_{0}(V_{xy})+\cos(k_{z})+\cos(k_{w}) ]$. By numerically diagonalizing the Floquet Hamiltonian $H_{F}$, we show the bulk gap as a function of $m$ and $V_{xy}$ in Fig.~\ref{fig5}(a), where the white region signifies the bulk gap close to zero. Similar to Sec.~\ref{SecIII.B}, we derive the effective Hamiltonian $H_{\rm{eff}}$ for this system based on the Floquet theory in the high-frequency limit (see Appendix~\ref{AppendixB} for details). By analytically solving for the effective Hamiltonian $H_{\rm{eff}}$, we can obtain the black dashed lines in Fig.~\ref{fig5}(a). The black dashed lines correspond to points where the bulk gap is zero, given by $m=0$, $m=\pm2$, $m=\pm 2\mathcal{J}_{0}(V_{xy})$, and $m=\pm2\pm 2\mathcal{J}_{0}(V_{xy})$. These black dashed lines fit well with the numerical calculations for the Floquet Hamiltonian.

In Fig.~\ref{fig5}(b), we show phase diagram in the ($m, V_{xy}$) plane. The color map corresponds to value of the second Chern number. Notably, the position of the phase transition point coincides with that of the closure point in the bulk gap. In the range of $1<|m|<2$, it can be observed that as the amplitude of the time-periodic driving $V_{xy}$ increases, the system undergoes a phase transition from a topological phase with $C_{2}=-3$ to another topological phase with $C_{2}=-1$ as shown in Fig.~\ref{fig5}(c). The quasienergy spectra of the quasi-3D system are shown in Figs.~\ref{fig6}(a) and \ref{fig6}(b). It is evident that the time-periodic driving $V_{xy}$ induces the emergence of the topological phase with $C_{2}=-1$, distinguished by a single gapless Dirac point. When $V_{xy}=1.5$ ($V_{xy}=3.5$), this Dirac point is located at the point ${\rm{Y}}~(k_{y}=\pi, k_{z}=0, k_{w}=0)$ [${\rm{G}}~(0, 0, 0)$].

Moreover, it is intriguing that for $|m|<1$, the periodic driving can induce the emergence of a topologically nontrivial system with an even second Chern number $C_{2}=\pm2$. Figure~\ref{fig5}(d) illustrates the second Chern number $C_{2}$ as a function of the amplitude $V_{xy}$ when $m=-0.4$. When the amplitude $V_{xy}$ exceeds the critical value $V_{xy,c}=0.9184$, the system transitions from the topological phase with $C_{2}=-3$ to another topological phase with $C_{2}=-2$. Within the interval $0.9184<V_{xy}<2.0415$, $C_{2}$ keeps a quantized plateau with $C_{2}=-2$. As shown in Fig.~\ref{fig6}(c), there are two Dirac points in the FBZ of the quasi-3D system at points ${\rm{Z}}~(k_{y}=0, k_{z}=\pi, k_{w}=0)$ and ${\rm{W}}~(k_{y}=0, k_{z}=0, k_{w}=\pi)$. When $V_{xy}>2.0415$, the topological properties of the system are destroyed until $V_{xy}>2.8371$. In the interval $V_{xy}\in(2.8371, 4.9307)$, the topologically nontrivial phase with the second Chern number $C_{2}=-2$ reemerges. In this case, there are two Dirac points in the FBZ as shown in Fig.~\ref{fig6}(d). It is noteworthy that in these topologically nontrivial systems induced by $V_{xy}$, when the OBC along the $z$ or $w$ direction, the value of the second Chern number does not coincide with the number of Dirac points (see Appendix~\ref{AppendixC} for details).

\section{Conclusion}
\label{Conclusion}
In this paper, we investigate the effects of high-frequency time-periodic driving in a 4D TI. First, we find that the time-periodic driving ${\boldsymbol{V}}(V_{x}, 0, 0, 0)$ can modulate the topological phase of the 4D TI when $V_{x}\neq0$. However, there is no newly formed topological phase with distinct second Chern number. The Floquet system exhibits similar results when $V=V_{n}\neq0$ ($n=x,y,z,w$) (see Appendix~\ref{AppendixE} for more details). In addition, we show that the time-periodic driving ${\boldsymbol{V}}(V_{x}, V_{y}, 0, 0)$ can force the system to convert from a topological phase with $C_{2}=\pm3$ to another topological phase with $C_{2}=\pm1$. When the Dirac mass $|m|<1$, the periodic driving $V_{xy}$ can additionally induce a topological phase characterized by an even second Chern number $C_{2}=\pm2$. Note that there is no such topologically nontrivial phase with an even second Chern number in the 4D static system. Similarly, the time-periodic driving ${\boldsymbol{V}}(V_{x}, V_{y}, V_{z}, 0)$ can also induce a topologically nontrivial phase with $C_{2}=\pm2$ (see Appendix~\ref{AppendixD}). Furthermore, the Floquet phase diagram can be explained by the approximation theory in the high-frequency limit.

Experimentally, electronic circuits can be used to realize the 4D system~\cite{10.1093/nsr/nwaa065, 10.1038/s41467-020-15940-3}. With tunable complex-phase elements, momentum components of the 4D Dirac model can be periodically driven to realize the time-dependent Hamiltonian~\cite{10.1038/s41467-023-36359-6}. Therefore, we expect that the time-periodically driven 4D TIs can be experimentally realized through topolectrical circuit networks.

It is worth mentioning that in our another work, we find that the time-periodic driving can induce a topological phase transition from a 4D normal insulator to a 4D FTI with nonzero second Chern number. In that work, the frequency of the time-periodic driving is smaller than the bandwidth of the static system, so the driven bands and the undriven bands overlap in the resonant quasienergy region. We consider two types of time-periodic driving, including a time-periodic onsite potential and a time-periodic vector potential. It is found that both types of the time-periodic driving can open the resonant quasienergy gap, and induce gapless topological boundary states.

\section*{Acknowledgments}
B.Z. was supported by the NSFC (Grant No. 12074107), the program of outstanding young and middle-aged scientific and technological innovation team of colleges and universities in Hubei Province (Grant No. T2020001) and the innovation group project of the Natural Science Foundation of Hubei Province of China (Grant No. 2022CFA012). R.C. acknowledges the support of NSFC (under Grant No. 12304195) and the Chutian Scholars Program in Hubei Province. Z.-R.L. was supported by the National Funded Postdoctoral Researcher Program (under Grant No. GZC20230751) and the Postdoctoral Innovation Research Program in Hubei Province (under Grant No. 351342).

\appendix

\section{Effective Hamiltonian for $V_{x}\neq0$}
\label{AppendixA}
By employing the Floquet theory in the high-frequency limit (i.e., $\omega\gg E_{W}$), the time-dependent system can be described by a time-independent effective Hamiltonian as~\cite{PhysRevLett.91.110404, PhysRevX.4.031027, PhysRevB.25.6622, PhysRevA.38.1739, PhysRevA.68.013820, Eckardt_2015, Bukov_2015}
\begin{align}
H_{\rm{eff}}=H_{0}+\sum_{l\neq0}\frac{[H_{-l},H_{+l}]}{l\omega}+\mathcal{O}(\omega^{-2}),
\end{align}
where $\omega$ is the frequency of the time-periodic driving. When $V_{x}\neq0$, $[H_{-l},H_{+l}]=0$, then
\begin{align}
H_{\rm{eff}}=H_{0},\label{A2}
\end{align}
where the specific form of $H_{0}$ is shown in Eq.~(\ref{Eq10}). In the static system, the bulk gap is closed at $m=\pm4, m=\pm2, m=0$. In these cases, the high symmetry points of the bulk gap closure in the FBZ are as follows:
\begin{align}
m=-4:~&\boldsymbol{\Gamma}~(k_{x}=0, k_{y}=0, k_{z}=0, k_{w}=0),\nonumber\\
m=-2:~&\textbf{X}~(\pi,0,0,0), \textbf{Y}~(0,\pi,0,0), \textbf{Z}~(0,0,\pi,0),\nonumber\\ &\textbf{W}~(0,0,0,\pi),\nonumber\\
m=0:~&\textbf{M}_{xy}~(\pi,\pi,0,0), \textbf{M}_{xz}~(\pi,0,\pi,0), \textbf{M}_{xw}~(\pi,0,0,\pi),\nonumber\\&\textbf{M}_{yz}~(0,\pi,\pi,0), \textbf{M}_{yw}~(0,\pi,0,\pi), \textbf{M}_{zw}~(0,0,\pi,\pi),\nonumber\\
m=2:~&\textbf{R}_{xyz}~(\pi,\pi,\pi,0), \textbf{R}_{xyw}~(\pi,\pi,0,\pi),\nonumber\\ &\textbf{R}_{xzw}~(\pi,0,\pi,\pi), \textbf{R}_{yzw}~(0,\pi,\pi,\pi),\nonumber\\
m=4:~&\textbf{Q}~(\pi,\pi,\pi,\pi).
\end{align}
When $V_{x}\neq0$, by solving for the effective Hamiltonian $H_{\rm{eff}}$ [Eq.~(\ref{A2})] at different high symmetry $\textbf{k}$ points, we obtain the equations for the Dirac mass $m$ versus $V_{x}$ when the bulk gap vanishes, $m=\pm3\pm\mathcal{J}_{0}(V_{x})$, $m=\pm1\pm\mathcal{J}_{0}(V_{x})$.

\section{Effective Hamiltonian for $V_{x}=V_{y}=V_{xy}\neq0$}
\label{AppendixB}
Similarly, by employing the Floquet theory in the high-frequency limit, the effective Hamiltonian for $V_{x}=V_{y}=V_{xy}\neq0$ is given by the following formula:
\begin{align}
H_{\rm{eff}}=H_{0}+\sum_{l\neq0}\frac{[H_{-l},H_{+l}]}{l\omega}+\mathcal{O}(\omega^{-2}),
\end{align}
where $[H_{-l},H_{+l}]=0$. Therefore,
\begin{align}
H_{\rm{eff}}=H_{0}=&\sin(k_{x})\mathcal{J}_{0}(V_{xy})\Gamma_{2}+\sin(k_{y})\mathcal{J}_{0}(V_{xy})\Gamma_{3}\nonumber\\
&+\sin(k_{z})\Gamma_{4}+\sin(k_{w})\Gamma_{5}+m(\textbf{k})\Gamma_{1},\label{B2}
\end{align}
where $m(\textbf{k})=m+c[\cos(k_{x})\mathcal{J}_{0}(V_{xy})+\cos(k_{y})\mathcal{J}_{0}(V_{xy})+\cos(k_{z})+\cos(k_{w}) ]$. By solving for the effective Hamiltonian $H_{\rm{eff}}$ [Eq.~(\ref{B2})], we conclude that the bulk gap vanishes when $m=0$, $m=\pm2$, $m=\pm 2\mathcal{J}_{0}(V_{xy})$, or $m=\pm2\pm 2\mathcal{J}_{0}(V_{xy})$.

\section{Quasienergy gap distributions for quasi-3D systems}
\label{AppendixC}
\begin{figure}[t]
	\includegraphics[width=8.5cm]{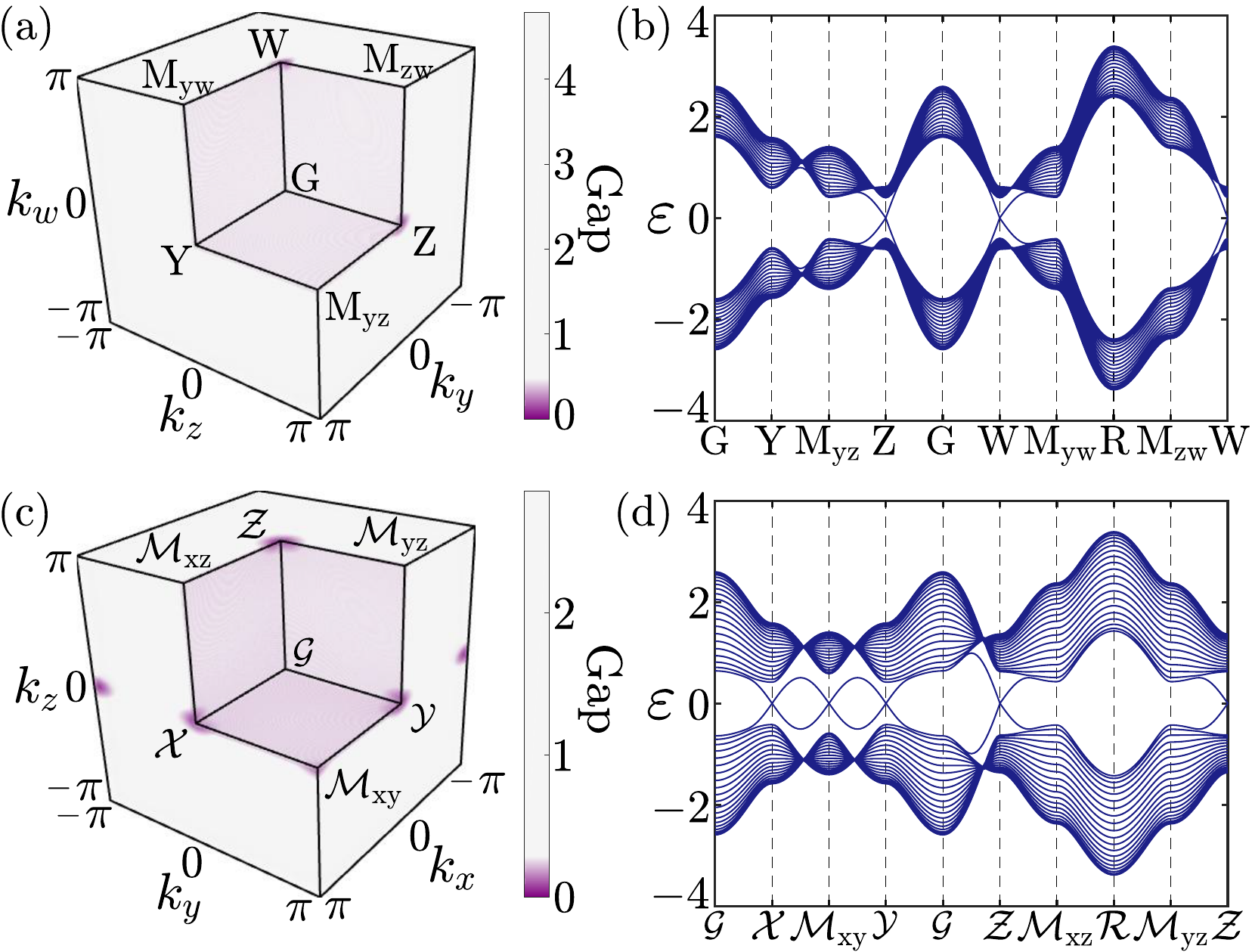} \caption{Quasienergy gap and spectra. (a) Quasienergy gap in the $(k_{y}, k_{z}, k_{w})$ space.  (b) Quasienergy spectra of the quasi-3D system. The horizontal axis shows the high symmetry points in the FBZ, ${\rm{G}}=(k_{y}=0, k_{z}=0, k_{w}=0)$, $\rm{Y}=(\pi, 0, 0)$, $\rm{Z}=(0, \pi, 0)$, $\rm{W}=(0, 0, \pi)$, $\rm{M_{yz}}=(\pi, \pi, 0)$, $\rm{M_{yw}}=(\pi, 0, \pi)$, $\rm{M_{zw}}=(0, \pi, \pi)$, $\rm{R}=(\pi, \pi, \pi)$. In (a) and (b), we take the OBC along the $x$ direction. (c) Quasienergy gap in the $(k_{x}, k_{y}, k_{z})$ space. (d) Quasienergy spectra of the quasi-3D system. The horizontal axis shows the high symmetry points in the FBZ, $\mathcal{G}=(k_{x}=0, k_{y}=0, k_{z}=0)$, $\mathcal{X}=(\pi, 0, 0)$, $\mathcal{Y}=(0, \pi, 0)$, $\mathcal{Z}=(0, 0, \pi)$, $\mathcal{M}_{\rm{xy}}=(\pi, \pi, 0)$, $\mathcal{M}_{\rm{xz}}=(\pi, 0, \pi)$, $\mathcal{M}_{\rm{yz}}=(0, \pi, \pi)$, $\mathcal{R}=(\pi, \pi, \pi)$. In (c) and (d), we take the OBC along the $w$ direction. For simplicity, here we only present the quasienergy spectra within the interval $\varepsilon\in[-\omega/2,\omega/2]$. In (a) and (c), the Dirac points where quasienergy gap vanishes are marked with the deepest purple color. Here, we choose $\omega=20$, $m=-0.4$, $V_{xy}=1.5$, and $L_{x}(L_{w})=20$.}%
	\label{figC1}
\end{figure}

In the main text, we note that in the static system or the Floquet system with $V_{x}\neq0$, the number of gapless 3D topological boundary states is equal to the value of the second Chern number. When $V_{x}=V_{y}=V_{xy}\neq0$, the periodic driving can induce the emergence of topological nontrivial system with $C_{2}=-2$.

In this appendix, we fix $m=-0.4$ and $V_{xy}=1.5$. The system is characterized by $C_{2}=-2$. In Figs.~\ref{figC1}(a) and \ref{figC1}(b), we show the quasienergy gap and spectra in the $(k_{y}, k_{z}, k_{w})$ space with OBC along the $x$ direction. In Fig.~\ref{figC1}(a), the color approaching purple indicate that the quasienergy gap is close to zero. It can be found that there are two gapless Dirac points at ${\rm{Z}}~(k_{y}=0, k_{z}=\pi, k_{w}=0)$ and ${\rm{W}}~(0, 0, \pi)$. This matches the value of the second Chern number. The number of gapless Dirac points remains the same for the system with OBC along the $y$ direction. However, there are four gapless Dirac points in the $(k_{x}, k_{y}, k_{z})$ space with OBC along the $w$ direction as shown in Figs.~\ref{figC1}(c) and \ref{figC1}(d). The results with OBC along the $z$ direction are the same as those along the $w$ direction. It is obvious that the number of gapless Dirac points does not match the value of the second Chern number for the system with OBC along the $z$ direction or the $w$ direction. When $m = -1.7$ and $V_{xy}=1.5$, the topologically nontrivial system with $C_{2}=-1$ exhibits the same phenomenon.

\section{Time-periodic driving ${\boldsymbol{V}}(V_{x}, V_{y}, V_{z}, 0)$}
\label{AppendixD}
\begin{figure}[t]
	\includegraphics[width=8.5cm]{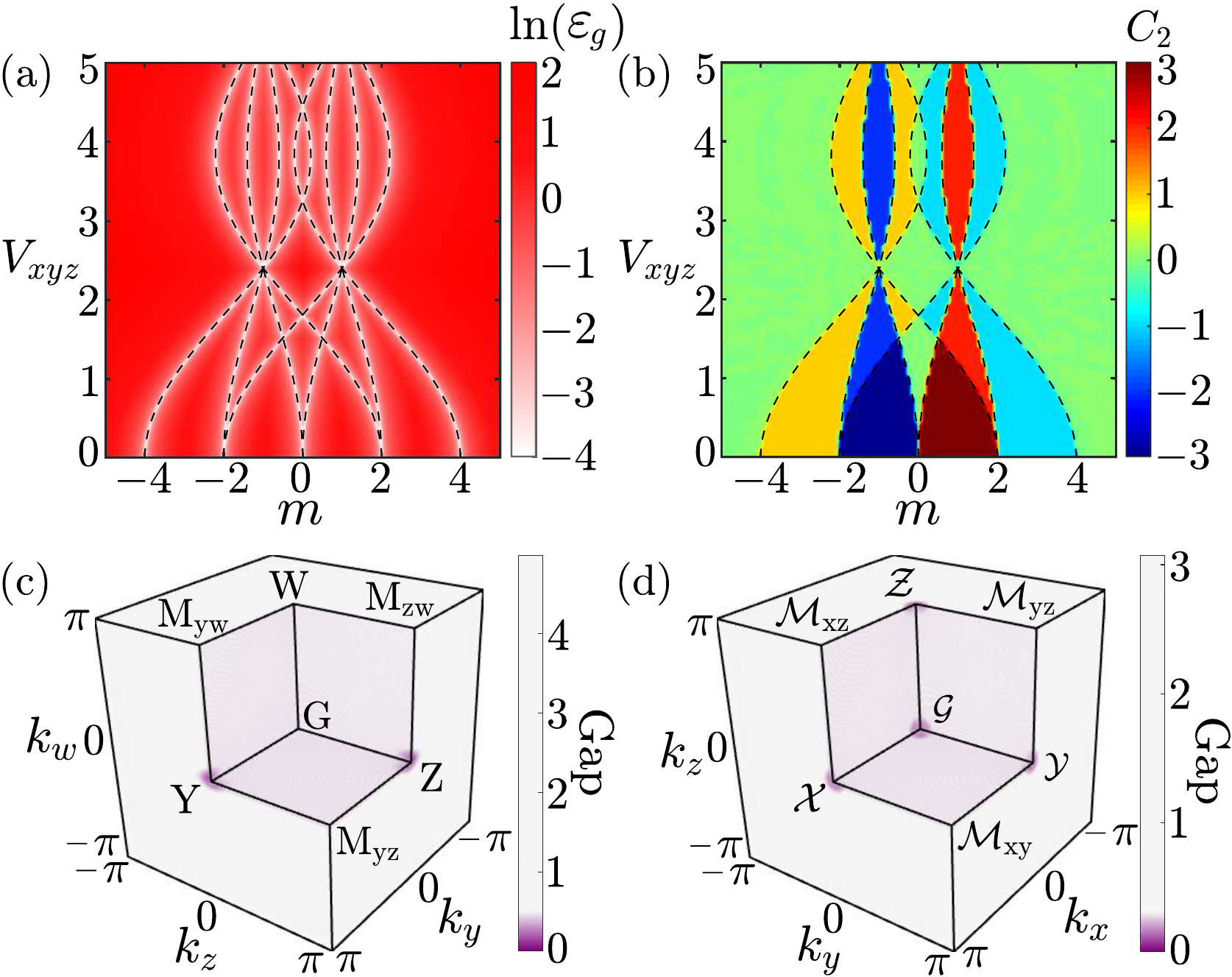} \caption{(a) The logarithm of the bulk gap as a function of $m$ and $V_{xyz}$. The white area indicates that the bulk gap is close to zero, and black dashed lines are given by $m=\pm1\pm 3\mathcal{J}_{0}(V_{xyz})$ and $m=\pm1\pm \mathcal{J}_{0}(V_{xyz})$. (b) Second Chern number $C_{2}$ as a function of $m$ and $V_{xyz}$. (c) Quasienergy gap in the $(k_{y}, k_{z}, k_{w})$ space with OBC along $x$ direction. (d) Quasienergy gap in the $(k_{x}, k_{y}, k_{z})$ space with OBC along $w$ direction. The Dirac points where quasienergy gap vanishes are marked with the deepest purple color. In (c) and (d), we set parameters $m=-1$, $V_{xyz}=1.5$, and $L_{x}(L_{w})=20$. Here, we choose $\omega=20$.}%
	\label{figD1}
\end{figure}

In this appendix, we consider the periodic driving ${\boldsymbol{V}}(V_{x}, V_{y}, V_{z}, 0)$, where $V_{x}=V_{y}=V_{z}=V_{xyz}\neq0$. Then, the diagonal block matrices in the Floquet Hamiltonian $H_{F}$ are given by:
\begin{align}
H_{0}=&\sin(k_{x})\mathcal{J}_{0}(V_{xyz})\Gamma_{2}+\sin(k_{y})\mathcal{J}_{0}(V_{xyz})\Gamma_{3}\nonumber\\
&+\sin(k_{z})\mathcal{J}_{0}(V_{xyz})\Gamma_{4}+\sin(k_{w})\Gamma_{5}+m(\textbf{k})\Gamma_{1},
\end{align}
and the off-diagonal block matrices
\begin{align}
H_{-l}=&\sum_{j=1}^{5}h_{-l, j}\Gamma_{j}, H_{+l}=H_{-l}^{\dagger},
\end{align}
with
\begin{align}
h_{-l, 1}=&\frac{c}{2}\sum_{n=x,y,z}[e^{i k_{n}}+(-1)^{l}e^{-i k_{n}}]\mathcal{J}_{l}(V_{xyz}),\nonumber\\
h_{-l, 2}=&-\frac{i}{2}[e^{i k_{x}}-(-1)^{l}e^{-i k_{x}}]\mathcal{J}_{l}(V_{xyz}),\nonumber\\
h_{-l, 3}=&-\frac{i}{2}[e^{i k_{y}}-(-1)^{l}e^{-i k_{y}}]\mathcal{J}_{l}(V_{xyz}),\nonumber\\
h_{-l, 4}=&-\frac{i}{2}[e^{i k_{z}}-(-1)^{l}e^{-i k_{z}}]\mathcal{J}_{l}(V_{xyz}),\nonumber\\
h_{-l, 5}=&0,
\end{align}
where $m(\textbf{k})=m+c[\cos(k_{x})\mathcal{J}_{0}(V_{xyz})+\cos(k_{y})\mathcal{J}_{0}(V_{xyz})+\cos(k_{z})\mathcal{J}_{0}(V_{xyz})+\cos(k_{w}) ]$. By solving the Floquet Hamiltonian, we show the bulk gap as a function of $m$ and $V_{xyz}$ in Fig.~\ref{figD1}(a), where the white region indicates the bulk gap approaching zero. Furthermore, in the high-frequency limit, we can find the effective Hamiltonian for $V_{x}=V_{y}=V_{z}=V_{xyz}\neq 0$:
\begin{align}
H_{\rm{eff}}=H_{0}+\sum_{l\neq0}\frac{[H_{-l},H_{+l}]}{l\omega}+\mathcal{O}(\omega^{-2}),
\end{align}
where $[H_{-l},H_{+l}]=0$. Therefore, we have
\begin{align}
H_{\rm{eff}}=&\sin(k_{x})\mathcal{J}_{0}(V_{xyz})\Gamma_{2}+\sin(k_{y})\mathcal{J}_{0}(V_{xyz})\Gamma_{3}\nonumber\\
&+\sin(k_{z})\mathcal{J}_{0}(V_{xyz})\Gamma_{4}+\sin(k_{w})\Gamma_{5}+m(\textbf{k})\Gamma_{1}.\label{D4}
\end{align}
By solving for the effective Hamiltonian $H_{\rm{eff}}$ [Eq.~(\ref{D4})], we conclude that the bulk gap vanishes when $m=\pm1\pm 3\mathcal{J}_{0}(V_{xyz})$ or $m=\pm1\pm \mathcal{J}_{0}(V_{xyz})$. In Fig.~\ref{figD1}, the black dashed lines represent the points where the bulk gap vanishes.

Correspondingly, we plot the phase diagram with respect to $m$ and $V_{xyz}$ in Fig.~\ref{figD1}(b). The color map represents the values of the second Chern number. When $|m|>2$, the periodic driving can destroy the topologically nontrivial properties of the system, leading to a reduction of the second Chern number of the system from $C_{2}=\pm1$ to $C_{2}=0$. Furthermore, within the interval $|m|<2$, the periodic driving can induce topological phase transitions, causing a transformation from a topological phase with $C_{2}=\pm 3$ to other topological phases with $C_{2}=\pm1$ or $C_{2}=\pm2$. Next, we fix $m=-1$ and $V_{xyz}=1.5$. The system is characterized by $C_{2}=-2$. Figures \ref{figD1}(c) and \ref{figD1}(d) respectively show the distribution of quasienergy gap for the systems with OBC along the $x$ and $w$ directions. There are two gapless Dirac points in $(k_{y}, k_{z}, k_{w})$ space with OBC along the $x$ direction. Similarly, there are two gapless Dirac points in quasi-3D systems with OBC along the $y$ or $z$ direction. However, there are four gapless Dirac points in the system with OBC along the $w$ direction, which does not match the value of the second Chern number $C_{2}=-2$.

\section{Time-periodic driving ${\boldsymbol{V}}(V_{x}, V_{y}, V_{z}, V_{w})$}
\label{AppendixE}
\begin{figure}[t]
	\includegraphics[width=8.5cm]{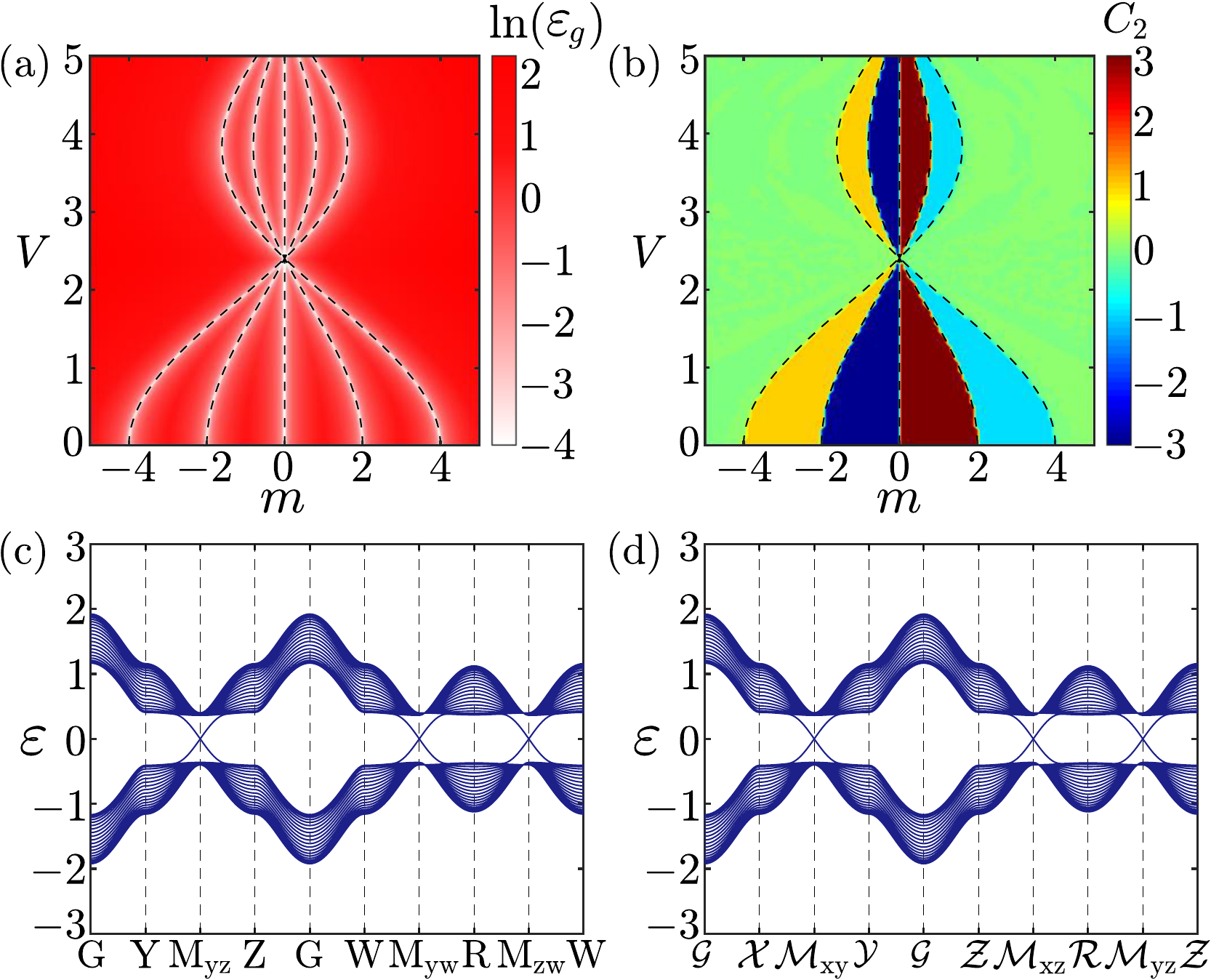} \caption{(a) The logarithm of the bulk gap as a function of $m$ and $V$. The white area indicates that the bulk gap is close to zero, and black dashed lines are given by $m=0$, $m=\pm2\mathcal{J}_{0}(V)$, and $m=\pm4\mathcal{J}_{0}(V)$. (b) Second Chern number $C_{2}$ as a function of $m$ and $V$. (c) Quasienergy spectra of the quasi-3D system with OBC along $x$ direction. The horizontal axis shows the high symmetry points in the FBZ, ${\rm{G}}=(k_{y}=0, k_{z}=0, k_{w}=0)$, $\rm{Y}=(\pi, 0, 0)$, $\rm{Z}=(0, \pi, 0)$, $\rm{W}=(0, 0, \pi)$, $\rm{M_{yz}}=(\pi, \pi, 0)$, $\rm{M_{yw}}=(\pi, 0, \pi)$, $\rm{M_{zw}}=(0, \pi, \pi)$, $\rm{R}=(\pi, \pi, \pi)$. (d) Quasienergy spectra of the quasi-3D system with OBC along $w$ direction. The horizontal axis shows the high symmetry points in the FBZ, $\mathcal{G}=(k_{x}=0, k_{y}=0, k_{z}=0)$, $\mathcal{X}=(\pi, 0, 0)$, $\mathcal{Y}=(0, \pi, 0)$, $\mathcal{Z}=(0, 0, \pi)$, $\mathcal{M}_{\rm{xy}}=(\pi, \pi, 0)$, $\mathcal{M}_{\rm{xz}}=(\pi, 0, \pi)$, $\mathcal{M}_{\rm{yz}}=(0, \pi, \pi)$, $\mathcal{R}=(\pi, \pi, \pi)$. In (c) and (d), we set parameters $m=-0.4$, $V=3.5$, and $L_{x}(L_{w})=20$. Here, we choose $\omega=20$.}%
	\label{figE1}
\end{figure}

In this appendix, we consider the periodic driving ${\boldsymbol{V}}(V_{x}, V_{y}, V_{z}, V_{w})$, where $V_{x}=V_{y}=V_{z}=V_{w}=V\neq0$. Then, the diagonal block matrices in the Floquet Hamiltonian $H_{F}$ are given by:
\begin{align}
H_{0}=&[\sin(k_{x})\Gamma_{2}+\sin(k_{y})\Gamma_{3}+\sin(k_{z})\Gamma_{4}\nonumber\\
&+\sin(k_{w})\Gamma_{5}]\mathcal{J}_{0}(V)+m(\textbf{k})\Gamma_{1},
\end{align}
and the off-diagonal block matrices
\begin{align}
H_{-l}=&\sum_{j=1}^{5}h_{-l, j}\Gamma_{j}, H_{+l}=H_{-l}^{\dagger},
\end{align}
with
\begin{align}
h_{-l, 1}=&\frac{c}{2}\sum_{n=x,y,z,w}[e^{i k_{n}}+(-1)^{l}e^{-i k_{n}}]\mathcal{J}_{l}(V),\nonumber\\
h_{-l, 2}=&-\frac{i}{2}[e^{i k_{x}}-(-1)^{l}e^{-i k_{x}}]\mathcal{J}_{l}(V),\nonumber\\
h_{-l, 3}=&-\frac{i}{2}[e^{i k_{y}}-(-1)^{l}e^{-i k_{y}}]\mathcal{J}_{l}(V),\nonumber\\
h_{-l, 4}=&-\frac{i}{2}[e^{i k_{z}}-(-1)^{l}e^{-i k_{z}}]\mathcal{J}_{l}(V),\nonumber\\
h_{-l, 5}=&-\frac{i}{2}[e^{i k_{w}}-(-1)^{l}e^{-i k_{w}}]\mathcal{J}_{l}(V),
\end{align}
where $m(\textbf{k})=m+c[\cos(k_{x})+\cos(k_{y})+\cos(k_{z})+\cos(k_{w}) ]\mathcal{J}_{0}(V)$. Meanwhile, by employing the Floquet theory in the high-frequency limit, we can derive the effective Hamiltonian $H_{\rm{eff}}$ as follows:
\begin{align}
H_{\rm{eff}}=H_{0}+\sum_{l\neq0}\frac{[H_{-l},H_{+l}]}{l\omega}+\mathcal{O}(\omega^{-2}).
\end{align}
Since $[H_{-l},H_{+l}]=0$, therefore
\begin{align}
H_{\rm{eff}}=H_{0}=&[\sin(k_{x})\Gamma_{2}+\sin(k_{y})\Gamma_{3}+\sin(k_{z})\Gamma_{4}\nonumber\\
&+\sin(k_{w})\Gamma_{5}]\mathcal{J}_{0}(V)+m(\textbf{k})\Gamma_{1}.\label{E4}
\end{align}

Figures~\ref{figE1}(a) and \ref{figE1}(b) show the bulk gap and the second Chern number for the Floquet Hamiltonian in the ($m, V$) plane. In Figs.~\ref{figE1}(a) and \ref{figE1}(b), the black dashed lines [$m=0$, $m=\pm2\mathcal{J}_{0}(V)$, $m=\pm4\mathcal{J}_{0}(V)$] represent the points where the bulk gap vanishes, which are obtained by solving for the effective Hamiltonian $H_{\rm{eff}}$. As shown in Fig.~\ref{figE1}(b), the topological nontrivial regions gradually diminish with increasing the amplitude of the periodic driving, until they completely vanish at $V_{c}\approx2.405$. Then, as the periodic driving further intensifies, the topologically nontrivial phases reemerge. When $m=-0.4$ and $V=3.5$, the second Chern number of the Floquet system is $C_{2}=-3$. In Figs.~\ref{figE1}(c) and \ref{figE1}(d), we show the quasienergy spectra for the system with OBC along the $x$ and $w$ directions, respectively. One can observe that the number of gapless Dirac points is the same in the quasi-3D system with OBC along the $x$ or $w$ direction. In addition, there are also three gapless Dirac points in quasi-3D systems with OBC along the $y$ or $z$ direction.

\newpage
\bibliographystyle{apsrev4-1-etal-title_6authors}
%\bibliography{references}
%

\end{document}